\documentclass[runningheads]{llncs}
\usepackage[T1]{fontenc}
\usepackage{graphicx}
\usepackage{booktabs}
\usepackage[misc]{ifsym}

\usepackage{amsmath,amssymb,amsfonts}
\usepackage{graphicx}
\usepackage{textcomp}
\usepackage{xcolor}

\usepackage[ruled, vlined, nofillcomment, linesnumbered]{algorithm2e}
\usepackage{url}
\usepackage{graphicx}
\usepackage{subcaption}
\usepackage{booktabs}
\usepackage{verbatim}
\usepackage{float}
\usepackage{xspace}
\usepackage{tikz}
\usepackage{pgfplots}
\usepackage{pgfplotstable}

\usetikzlibrary{fit,external,positioning}
\usetikzlibrary{shapes.geometric}

\usepackage[numbers,sort&compress]{natbib}
\usepackage{etoolbox}
\makeatletter
\patchcmd{\@algocf@start}
  {-1.5em}
  {0pt}
  {}{}
\makeatother

\renewcommand{\refname}{References}
\makeatletter
\renewcommand\bibsection{%
  \section*{{\refname}\@mkboth{\refname}{\refname}}%
}%
\makeatother

\newcommand{\set}[1]{\left\{#1\right\}}

\newcommand{\fpr}[1]{\mathopen{}\left(#1\right)}

\newcommand{\abs}[1]{{\left|#1\right|}}

\newcommand{\np}{\textbf{NP}}
\newcommand{\zpp}{\textbf{ZPP}}

\newcommand{\define}{\leftarrow}

\DeclareRobustCommand{\dispfunc}[2]{%
  \ensuremath{%
  \ifthenelse{\equal{#2}{}}%
    {\mathit{#1}}%
    {\mathit{#1}\fpr{#2}}}}

\newcommand{\dens}[1]{\dispfunc{d}{#1}}

\newcommand{\diff}[1]{\dispfunc{\Delta}{#1}}

\newcommand{\bigO}[1]{\dispfunc{\mathcal{O}}{#1}}

\newcommand{\problemdts}{\textsc{TDS}\xspace}
\newcommand{\problemdcs}{\textsc{MDS}\xspace}

\newcommand{\problemcdcsm}{\textsc{FDS}\xspace}
\newcommand{\problemcdcsdiff}{\textsc{SDS}\xspace}

\newcommand{\alggrd}{\textsc{SDS-Grd}\xspace}
\newcommand{\alggrdfms}{\textsc{FDS-Grd}\xspace}

\newcommand{\algipcm}{\textsc{FDS-IP}\xspace}
\newcommand{\algipdcs}{\textsc{MDS-IP}\xspace}
\newcommand{\algipdiff}{\textsc{SDS-IP}\xspace}
\newcommand{\algexact}{\textsc{IP}\xspace}
\newcommand{\alggrdshort}{\textsc{GR}\xspace}

\newcommand{\dentds}{\ensuremath{d_{\mathit{tds}}}\xspace}
\newcommand{\denmds}{\ensuremath{d_{\mathit{mds}}}\xspace}

\newcommand{\sbsnorm}{\ensuremath{\sigma_{\mathit{nrm}}}\xspace}

\newcommand{\dtname}[1]{\textsl{#1}}

\newcommand{\calG}{\ensuremath{{\mathcal G}}}

\newcommand{\prbclique}{\textsc{Clique}\xspace}

\SetKwComment{tcpas}{\{}{\}}
\SetCommentSty{textnormal}
\SetArgSty{textnormal}
\SetKw{False}{false}
\SetKw{True}{true}
\SetKw{Null}{null}
\SetKwInOut{Output}{output}
\SetKwInOut{Input}{input}
\SetKw{AND}{and}
\SetKw{OR}{or}
\SetKw{Break}{break}

\SetKwFor{Until}{until}{do}{}

\definecolor{yafcolor1}{rgb}{0.4, 0.165, 0.553}
\definecolor{yafcolor2}{rgb}{0.949, 0.482, 0.216}
\definecolor{yafcolor3}{rgb}{0.47, 0.549, 0.306}
\definecolor{yafcolor4}{rgb}{0.925, 0.165, 0.224}
\definecolor{yafcolor5}{rgb}{0.141, 0.345, 0.643}
\definecolor{yafcolor6}{rgb}{0.965, 0.933, 0.267}
\definecolor{yafcolor7}{rgb}{0.627, 0.118, 0.165}
\definecolor{yafcolor8}{rgb}{0.878, 0.475, 0.686}
\definecolor{yafcolor9}{rgb}{0.178, 0.475, 0.686}
\definecolor{yafcolor10}{rgb}{0.878, 0.475, 0.286}

\SetKw{Output}{output}

\newcommand{\poly}{\textbf{P}}

\DeclareRobustCommand{\dispfunc}[2]{%
	\ensuremath{%
		\ifthenelse{\equal{#2}{}}%
			{\mathit{#1}}%
			{\mathit{#1}\fpr{#2}}}}

\SetKw{Break}{break}

\pgfdeclarelayer{background}
\pgfdeclarelayer{foreground}
\pgfsetlayers{background,main,foreground}

\definecolor{yafaxiscolor}{rgb}{0.3, 0.3, 0.3}

\definecolor{yafcolor1}{rgb}{0.4, 0.165, 0.553}
\definecolor{yafcolor2}{rgb}{0.949, 0.482, 0.216}
\definecolor{yafcolor3}{rgb}{0.47, 0.549, 0.306}
\definecolor{yafcolor4}{rgb}{0.925, 0.165, 0.224}
\definecolor{yafcolor5}{rgb}{0.141, 0.345, 0.643}
\definecolor{yafcolor6}{rgb}{0.965, 0.933, 0.267}
\definecolor{yafcolor7}{rgb}{0.627, 0.118, 0.165}
\definecolor{yafcolor8}{rgb}{0.878, 0.475, 0.686}

\newlength{\yafaxispad}
\newlength{\yaftlpad}
\newlength{\yaflabelpad}
\newlength{\yafaxiswidth}
\newlength{\yafticklen}
\makeatletter
\def\pgfplots@drawtickgridlines@INSTALLCLIP@onorientedsurf#1{}
\makeatother

\newcommand{\yafdrawyaxis}[2]{
	\pgfplotstransformcoordinatey{#1}\let\ymincoord=\pgfmathresult 
	\pgfplotstransformcoordinatey{#2}\let\ymaxcoord=\pgfmathresult 
	\pgfsetlinewidth{\yafaxiswidth} 
	\pgfsetcolor{yafaxiscolor}
	\pgfpathmoveto{\pgfpointadd{\pgfpointadd{\pgfplotspointrelaxisxy{0}{0}}{\pgfqpointxy{0}{\ymincoord}}}{\pgfqpoint{\yafaxispad}{-0.5\yafaxiswidth}}}
	\pgfpathlineto{\pgfpointadd{\pgfpointadd{\pgfplotspointrelaxisxy{0}{0}}{\pgfqpointxy{0}{\ymaxcoord}}}{\pgfqpoint{\yafaxispad}{0.5\yafaxiswidth}}}
	\pgfusepath{stroke}
}

\newcommand{\yafdrawaxis}[4]{
	\pgfplotstransformcoordinatex{#1}\let\xmincoord=\pgfmathresult 
	\pgfplotstransformcoordinatex{#2}\let\xmaxcoord=\pgfmathresult 
	\pgfplotstransformcoordinatey{#3}\let\ymincoord=\pgfmathresult 
	\pgfplotstransformcoordinatey{#4}\let\ymaxcoord=\pgfmathresult 
	\pgfsetlinewidth{\yafaxiswidth} 
	\pgfsetcolor{yafaxiscolor}
	\pgfpathmoveto{\pgfpointadd{\pgfpointadd{\pgfplotspointrelaxisxy{0}{0}}{\pgfqpointxy{\xmincoord}{0}}}{\pgfqpoint{-0.5\yafaxiswidth}{\yafaxispad}}}
	\pgfpathlineto{\pgfpointadd{\pgfpointadd{\pgfplotspointrelaxisxy{0}{0}}{\pgfqpointxy{\xmaxcoord}{0}}}{\pgfqpoint{0.5\yafaxiswidth}{\yafaxispad}}}
	\pgfpathmoveto{\pgfpointadd{\pgfpointadd{\pgfplotspointrelaxisxy{0}{0}}{\pgfqpointxy{0}{\ymincoord}}}{\pgfqpoint{\yafaxispad}{-0.5\yafaxiswidth}}}
	\pgfpathlineto{\pgfpointadd{\pgfpointadd{\pgfplotspointrelaxisxy{0}{0}}{\pgfqpointxy{0}{\ymaxcoord}}}{\pgfqpoint{\yafaxispad}{0.5\yafaxiswidth}}}
	\pgfusepath{stroke}
}

\pgfplotscreateplotcyclelist{yaf}{%
{yafcolor5,mark options={scale=0.75},mark=o}, 
{yafcolor2,mark options={scale=0.75},mark=square},
{yafcolor3,mark options={scale=0.75},mark=triangle},
{yafcolor4,mark options={scale=0.75},mark=o},
{yafcolor1,mark options={scale=0.75},mark=o},
{yafcolor6,mark options={scale=0.75},mark=o},
{yafcolor7,mark options={scale=0.75},mark=o},
{yafcolor8,mark options={scale=0.75},mark=o}} 

\pgfkeys{/pgf/number format/.cd,1000 sep={\,}}

\pgfplotsset{axis y line=left, axis x line=bottom,
	tick align=outside,
	tickwidth=\yafticklen,
	clip = false,
    x axis line style= {-, line width = 0pt, color=black!0},
    y axis line style= {-, line width = 0pt, color=black!0},
    x tick style= {line width = \yafaxiswidth, color=yafaxiscolor, yshift = \yafaxispad},
    y tick style= {line width = \yafaxiswidth, color=yafaxiscolor, xshift = \yafaxispad},
    x tick label style = {font=\small, yshift = \yaftlpad, inner xsep = 0pt},
    y tick label style = {font=\small, xshift = \yaftlpad},
    every axis y label/.style = {at = {(ticklabel cs:0.5)}, rotate=90, anchor=center, font=\small, yshift = -\yaflabelpad, inner sep = 0pt},
    every axis x label/.style = {at = {(ticklabel cs:0.5)}, anchor=center, font=\small, yshift = \yaflabelpad},
    x tick label style = {font=\small, yshift = 1pt},
    grid = major,
    major grid style  = {dash pattern = on 1pt off 3 pt},
	every axis plot post/.append style= {line width=\yafaxiswidth} ,
	legend cell align = left,
	legend style = {inner sep = 1pt, cells = {font=\scriptsize}},
	legend image code/.code={%
		\draw[mark repeat=2,mark phase=2,#1] 
		plot coordinates { (0cm,0cm) (0.15cm,0cm) (0.3cm,0cm) };%
	} 
}

\newcommand{\pgfprintduration}[1]{%
	\ifthenelse{\equal{#1}{}}{---}{%
	\pgfmathsetmacro{\minutes}{floor(#1 / 60)}%
	\pgfmathsetmacro{\seconds}{#1 - 60*\minutes}%
	\pgfmathifthenelse{\minutes > 0}{"\pgfmathprintnumber{\minutes}m \pgfmathprintnumber[fixed,precision=0]{\seconds}s"}{"\pgfmathprintnumber{\seconds}s"}\pgfmathresult}}

\begin{document}

\title{Fair densest subgraph across multiple graphs}
\toctitle{Fair densest subgraph across multiple graphs}

\author{Chamalee Wickrama Arachchi  \Letter  \and Nikolaj Tatti} 
\tocauthor{Chamalee Wickrama Arachchi, Nikolaj Tatti} 

\institute{HIIT, University of Helsinki, firstname.lastname@helsinki.fi}

\maketitle

\begin{abstract}
Many real-world networks can be modeled as graphs.
Finding dense subgraphs is a key problem in graph mining with applications in diverse domains.
In this paper, we consider two variants of the densest subgraph problem where multiple graph snapshots are given and the goal is to find a fair densest subgraph without over-representing the density among the graph snapshots.
More formally, given a set of graphs and input parameter $\alpha$, we find a dense subgraph maximizing the sum of densities across snapshots such that the difference between the maximum and minimum induced density is at most $\alpha$. We prove that this problem is \np-hard and present an integer programming based, exact algorithm and a practical polynomial-time heuristic.  We also consider a minimization variant where given an input parameter $\sigma$, we find a dense subgraph which minimizes the difference between the maximum and minimum density while inducing a total density of at least $\sigma$ across the graph snapshots. We prove the \np-hardness of the problem and propose two algorithms: an exponential time algorithm based on integer programming and a greedy algorithm.
We present an extensive experimental study that shows that our algorithms can find the ground truth in synthetic dataset and produce good results in real-world datasets. Finally, we present case studies that show the usefulness of our problem.

\end{abstract}

\section{Introduction}

The problem of dense subgraph discovery is an important tool in graph mining with applications in temporal pattern mining in financial markets~\cite{xiaoxi2009migration}, social network analysis~\cite{semertzidis2019finding}, and biological system analysis~\cite{fratkin2006motifcut}. 
On the other hand, multi-layer graph networks naturally exist in real-world, complex networks and have gained a significant amount of attention~\cite{jethava2015finding,semertzidis2019finding,rozenshtein2020finding,galimberti2020core,arachchi2023jaccard}.

Among many definitions of a dense component, the ratio between the number of induced edges and the number of nodes has been popular since Goldberg~\cite{goldberg1984finding}
proposed a polynomial-time, exact algorithm to find the densest subgraph of a single graph and Charikar~\cite{charikar2000greedy} proposed a greedy approximation algorithm.

Given a graph sequence, a natural way to find the densest subgraph is to find a common subgraph that maximizes the sum of densities in individual snapshots. This is done by first flattening the graphs into a single weighted graph and solving the weighted densest graph problem~\cite{semertzidis2019finding}.

While this approach finds the densest subgraph, it may be the case that the densities are unevenly distributed. That is, the sum is dominated by the density of a single or few snapshots while the densities in the remaining snapshots are unfairly low or even 0.

In this paper, we consider a fair variant of the densest subgraph problem, where we require that the densities are close to each other across the snapshots. 
The topic of fairness is relatively new within the data mining and machine learning community which has been studied with classical problems like clustering, community detection, and anomaly detection~\cite{ahmadian2019clustering,mehrabi2019debiasing, mehrabi2021survey,anagnostopoulos2020spectral}.
The fairness in the context of the densest subgraph problems is not well-studied yet.

More specifically, we consider two variants of a fair dense subgraph problem.
In our first problem, we find a fair subgraph that maximizes the sum of the densities while being evenly distributed.
To ensure that the total density is fairly distributed among the graph snapshots, we constrain the difference between the maximum and minimum density induced by the subgraph via an input parameter that specifies the maximum allowed density difference.
In our second problem, given a pre-defined, minimum value for the total density as an input parameter, our goal is to minimize the gap between the maximum and minimum induced density over the graph sequence.
 
We show that both of our problems are \np-hard, and that the second problem is inapproximable.
To solve our problems, we first propose two exact algorithms based on integer linear programming which run in exponential time. We also consider two heuristics that run in polynomial time in the size of the input graph.

The remainder of the paper is organized as follows.
In Section \ref{sec:prel}, we provide preliminary notation along with the
formal definitions of our optimization problems. Next, we prove
\np-hardness of our problems in Section~\ref{sec:compl}. All our
algorithms and their running times are presented in Section~\ref{sec:algorithm}. Related work is
discussed in Section~\ref{sec:related}. In Section~\ref{sec:exp} we present an extensive
experimental study both with synthetic and real-world datasets followed by a case study. 
Finally, Section~\ref{sec:conclusions} summarizes the paper and provides directions for future work.

\section{Preliminary notation and problem definition}\label{sec:prel}

We begin by providing preliminary notation and formally defining our problem.

Our input is a sequence of graphs $\calG = (G_1, \ldots, G_r)$, where each snapshot $G_i = (V, E_i)$
is defined over the same set of nodes. We often denote the number of nodes and edges by $n = \abs{V}$. 

Given a graph $G = (V, E)$, and a set of nodes $S \subseteq V$, we define $E(S) \subseteq E$ to be the subset of edges
having both endpoints in $S$. We also write $m(S) = \abs{E(S)}$. Given a  graph sequence $\calG = (G_1, \ldots, G_r)$ with $G_i = (V, E_i)$, we write $E(S, G_i) = E_i(S)$ and $m(S, G_i) = \abs{E(S, G_i)}$.

As mentioned before, our goal is to find dense subgraphs in a temporal network, and for that
we need to quantify the density of a subgraph. More formally,
assume an unweighted graph $G = (V, E)$, and let $S \subseteq V$.
We define the \emph{density} $\dens{S, G}$ of a graph $G_i$ induced by node set $S$,
and extend this definition for a sequence of graphs $\calG = (G_1, \ldots, G_k)$
as
\[
	\dens{S, G_i} = \frac{\abs{E(S, G_i)}}{\abs{S}}
	\quad\quad \text{and}\quad\quad
	\dens{S, \calG} = \sum_{i = 1}^r \dens{S, G_i}
	\quad.
\]

We first state the problem of finding a common subgraph in a graph sequence which maximizes the sum of the densities proposed by Semertzidis et al.~\cite{semertzidis2019finding}.

\begin{problem}[Total densest subgraph problem~(\problemdts)]
\label{pr:dts}
Given a graph sequence $\calG = (G_1,\dots,G_r)$, with  $G_i = (V, E_i)$, find a common subset of vertices
$S$, 
such that $\dens{S, \calG}$ is maximized.
\end{problem}

This problem can be solved by first flattening the graph sequence into one weighted graph, where an edge weight is the number of snapshots in which an edge occurs.
The problem is then a standard (weighted) densest subgraph problem that can be solved
using the exact method given by Goldberg~\cite{goldberg1984finding} in $\bigO{n(n + m) (\log n + \log r)}$ time.

Next, we introduce our main problem where an additional fairness constraint of the induced densities is considered. 
To this end, 
we denote the difference between the maximum and minimum induced density as
\[
    \diff{S, \calG} = \max_i \dens{S, G_i} - \min_i \dens{S, G_i} \quad.
\]
Given a sequence of graph snapshots and an input parameter $\alpha$, our goal
is to find a subset of vertices,
such that the sum of the densities of subgraphs is maximized while maintaining the difference $\diff{S, \calG}$ at most $\alpha$.

\begin{problem}[Fair densest subgraph problem~(\problemcdcsm)]
\label{pr:fair-dense-min}
Given a graph sequence $\calG = (G_1,\dots,G_r)$, with  $G_i = (V, E_i)$ and  real number  $\alpha$, find a  subset of vertices
$S$, 
such that $\dens{S, \calG}$ is maximized
and $\diff{S, \calG} \leq  \alpha$.
\end{problem}

Note that for $\alpha = 0$, the \problemcdcsm problem is equivalent to finding a subgraph which induces exactly the same density over each snapshot while maximizing the total density.
On the other hand, setting $\alpha = \infty$ reduces \problemcdcsm to \problemdts.




Next, we present a minimization variant of \problemcdcsm problem where given an input parameter $\sigma$, the goal is to find a subset of vertices $S$ that minimizes the difference $\diff{S, \calG}$  while inducing a total density of at least $\sigma$. 

\begin{problem}[The smallest difference densest subgraph problem~(\problemcdcsdiff)]
\label{pr:diff}
Given a graph sequence $\calG = (G_1,\dots,G_r)$, with  $G_i = (V, E_i)$ and real number $\sigma$, find a common subset of vertices $S$, 
such that the density induced by $S$  over  $\calG$  is at least $\sigma$ and $\diff{S, \calG}$ is minimized.
\end{problem}

Finally, we state the minimum densest subgraph problem~\cite{jethava2015finding} where the goal is to find a common subgraph which maximizes the minimum density.

\begin{problem}[Minimum densest subgraph~(\problemdcs)]
\label{pr:dcs}
Given a graph sequence $\calG = (G_1,\dots,G_r)$, with  $G_i = (V, E_i)$, find a common subset of vertices
$S$, 
such that $\min_i \dens{S, G_i}$, the minimum density induced by  $S$  over  $\calG$  is maximized.
\end{problem}

\section{Computational complexity}\label{sec:compl}

In this section, we show that both of our problems are \np-hard.
We saw in the previous section that if
we set $\alpha  = \infty$, then \problemcdcsm reduces to \problemdts, which can be solved exactly in
polynomial time. However, \problemcdcsm is \np-hard when $\alpha = 0$.

\begin{proposition} 
\label{prop:np}
\problemcdcsm is \np-hard.
\end{proposition} 
\begin{proof}
We prove the hardness from $k$-\prbclique, a problem where, given a graph $H$, we are asked if there is a clique of size at least $k$.

Assume that we are given a graph $H = (V, E)$ with $n$ nodes, $n \geq k$. We set $\alpha = 0$.
The graph snapshot $G_1$ consists of the
graph $H$ and an additional set of $k$ singleton vertices $U$. $G_2$ consists of a $k$-clique connecting the vertices in $U$. 

We claim that there is a subset $S$ yielding $\dens{S, \calG} =  (k - 1)/2$ if and only if there is an $k$-clique in $H$.

Assume that  there is a subset $S$ yielding $\dens{S, \calG} =  (k - 1)/2$.
Since the value of objective is $(k - 1)/2$, we have $\dens{S, G_1} = \dens{S, G_2} = (k - 1)/4$. 
Let $S = W \cup T$ where $W \subseteq V$ and denotes the subset of vertices from $H$ and $T \subseteq U$ is the subset of vertices from  $U$ in $S$. 

Assume that $\abs{W} < \abs{T}$. Since $\abs{T} \leq k$, $\abs{W} < k$.  The density induced on $G_1$ is bounded by $\dens{S, G_1} \leq \frac{{\abs{W} \choose 2}}{\abs{T} + \abs{W}} < \frac{{\abs{W} \choose 2}}{2\abs{W}} < \frac{k - 1}{4}$, which is a contradiction.
Assume that  $\abs{W} > \abs{T}$. Then the density induced on $G_2$ is bounded by $\dens{S, G_2} = \frac{{\abs{T} \choose 2}}{\abs{T} + \abs{W}} < \frac{{\abs{T} \choose 2}}{2\abs{T}} = \frac{\abs{T} - 1}{4} \leq \frac{k - 1}{4}$, again a contradiction. Therefore, $\abs{W} =  \abs{T}$. 

Consequently, $\dens{S, G_2} = (\abs{T} - 1)/4$, implying that $\abs{T} = k$. Finally, $\frac{k - 1}{4} = \dens{S, G_1} = \frac{\abs{E(S)}}{2k}$ implies that $\abs{E} = {k \choose 2}$, that is, $W$ is a $k$-clique in $H$.


On the other hand, assume there is a clique $C$ of size $k$ in $H$.
Set $S = C \cup U$. 
Immediately, $\dens{S, \calG} =   (k - 1)/2$ proving the claim.
\qed
\end{proof}

A similar proof will show that \problemcdcsdiff is \np-hard, and inapproximable.

\begin{proposition} 
\label{prop:np2}
\problemcdcsdiff is \np-hard.
Unless $\poly = \np$, there is no polynomial-time algorithm with multiplicative approximation guarantee for \problemcdcsdiff. 
\end{proposition} 

\begin{proof}
We use the same reduction from $k$-\prbclique as in the proof of Proposition~\ref{prop:np}.
We also set $\sigma = (k - 1)/2$. If there is a clique $C$ in $H$, then selecting $S = C \cup U$ yields $\diff{S, \calG} = 0$.
On the other hand, if $\diff{S, \calG} = 0$, then $\dens{S, G_1} = \dens{S, G_2} = (k - 1)/4$, and the argument in the proof of Proposition~\ref{prop:np} shows that there must be a $k$-clique in $H$.
In summary, the difference $\diff{S, \calG} = 0$ for a solution $S$ if and only if there is a $k$-clique in $H$.

This also immediately implies that there is no polynomial-time algorithm with multiplicative approximation guarantee since this algorithm can be then used to test whether there is a set $S$ with $\diff{S, \calG} = 0$.
\qed
\end{proof}
\section{Algorithms}\label{sec:algorithm}

In this section, we present our algorithms for \problemcdcsm and \problemcdcsdiff problems. We present two exact algorithms based on fractional programming~\citep{dinkelbach1967nonlinear} and integer linear programming. Both algorithms may require exponential time but in practice can provide a solution in reasonable time for moderately-sized graphs. In addition, we propose two polynomial time heuristics.

\subsection{Exact algorithm for \problemcdcsm}

We will present an integer programming based algorithm algorithm that finds an exact or near-optimal solution for \problemcdcsm problem. 
To this end, let us first define the \emph{edge} difference as
\[
    b(S, \calG) = \max_i m(S, G_i) - \min_i m(S, G_i)\quad.
\]
To solve \problemcdcsm we consider the following auxiliary problem.

\begin{problem}[$\problemcdcsm(\gamma)$]
\label{pr:label-subgrap-str-wk-alpha}
Given a graph sequence $\calG = (G_1,\dots,G_r)$, with  $G_i = (V, E_i)$, and two numbers $\alpha, \gamma$, find a subset of vertices $S \subseteq V$ maximizing  $ \sum_{i = 1}^r m(S, G_i) - \gamma \abs{S}$ such that $b(S, \calG) \leq  \alpha \abs{S}$.
\end{problem}

Next, we show the relationship between $\problemcdcsm(\gamma)$ and \problemcdcsm. This connection is an example of fractional programming~\citep{dinkelbach1967nonlinear}.

\begin{proposition}
\label{prop:frac}
Let $S(\gamma)$  be the subgraph  solving $\problemcdcsm(\gamma)$. Similarly, let $S^*$  be the optimal solution for \problemcdcsm. Write $\gamma^* =  \dens{S^*, \calG}$. If $\gamma > \gamma^*$, then $S(\gamma) = \emptyset$. If $\gamma < \gamma^*$, then $S(\gamma) \neq \emptyset$ and $\dens{S(\gamma),  \calG} > \gamma$.
\end{proposition}

\begin{proof}
Let $f(S) = \sum_{i = 1}^r m(S, G_i)$.
We know that
\begin{equation}
\label{eq:alphapos}
    f(S(\gamma)) - \gamma \abs{S(\gamma)} \geq f(\emptyset) - \gamma\abs{\emptyset} = 0.
\end{equation}

Let us first assume that $\gamma > \gamma^*$. If $S(\gamma) \neq \emptyset$, then Eq.~\ref{eq:alphapos} implies that $\dens{S(\gamma),  \calG} \geq \gamma > \gamma^*$, which contradicts the optimality of $\gamma^*$. Thus, $S(\gamma) = \emptyset$.

Next, let us assume that $\gamma < \gamma^*$. Then
\[
\begin{split}
    f(S(\gamma)) - \gamma \abs{S(\gamma)} & \geq f(S^*) - \gamma \abs{S^*}  > f(S^*) - \gamma^* \abs{S^*} = 0.
\end{split}
\]
Thus $f(S(\gamma)) >  \gamma \abs{S(\gamma)}$, and consequently $S(\gamma) \neq \emptyset$ and $\dens{S(\gamma), \calG} > \gamma$. \qed
\end{proof}

Next, we use Proposition~\ref{prop:frac} to solve our main problem \problemcdcsm. We find the (almost) largest $\gamma$ for which $\problemcdcsm(\gamma)$ yields a non-empty solution. Then $\problemcdcsm(\gamma)$ for such $\gamma$ yields (almost) optimal solution.

We can solve $\problemcdcsm(\gamma)$ with an integer linear program,

\begin{align}
	\textsc{maximize}&&  \sum_{k = 1}^r \sum_{ ij \in E_k} x_{ij} & - \gamma \sum_{i = 1}^{n} y_i \nonumber \\  
	\textsc{subject to}&& x_{ij} & \leq y_i &  ij \in E  \label{ip_con_1} \\
	&&x_{ij} & \leq y_j &  ij \in E \label{ip_con_2} \\
        &&x_{ij} & \geq y_i  + y_j - 1, &  ij \in E \label{ip_con_3}\\
	&&\sum_{ij \in E_k} x_{ij} - \sum_{ij \in E_\ell} x_{ij}  & \leq \alpha \sum_{i = 1}^{n} y_i   &  k, \ell = 1,\dots, r \label{ip_con_5}\\
	&&x_{ij},  y_j  & \in \{0,1\} \nonumber\quad.
\end{align}

To see why this program solves $\problemcdcsm(\gamma)$, 
let $S \subseteq V$  be a solution to our $\problemcdcsm(\gamma)$, 
The indicator variable $y_i$ denotes whether the node $i \in S$, and the indicator variable $x_{ij}$ denotes whether both endpoints of edge $ij$ are in $S$.
Note that Constraints~\ref{ip_con_1}--\ref{ip_con_3} force $x_{ij} = \min(y_i, y_j)$.

Furthermore, Constraint~\ref{ip_con_5} ensures that $b(S, \calG) \leq \alpha \abs{S}$. 

Proposition~\ref{prop:frac} allows to maximize $\gamma$
with a binary search. Here, we choose the initial interval $(L, U)$ by setting lower threshold $L = 0$ and upper threshold $U = r\frac{n - 1}{2}$, and keep halving the interval until $\abs{ U - L} \leq  \epsilon L$, where $\epsilon > 0$  is an input parameter. Finally, we return the solution of $\problemcdcsm(L)$ as the final solution to \problemcdcsm. We refer to this algorithm as \algipcm.
Next, we show that \algipcm yields $1/(1+\epsilon)$ approximation guarantee, or exact solution if $\epsilon$ is small enough.

\begin{proposition}
Assume a graph sequence $\calG = (G_1,\dots,G_r)$, $\alpha \geq 0$, and $\epsilon > 0$. 
Let $\gamma$ be the score of the solution returned by $\algipcm$ and let $\gamma^*$ be the optimal score of \problemcdcsm. Then $\gamma \geq \gamma^*/(1 + \epsilon)$. If $\epsilon \leq \frac{1}{rn^3}$, then $\gamma = \gamma^*$.
\label{prop:opt-ip-approx}
\end{proposition}

\begin{proof}
Let $L$ and $U$ be the values of the interval when binary search is terminated. Note that $\gamma \geq L$ due to Proposition~\ref{prop:frac}.
We know that $U - L \leq \epsilon L$ and $L \leq \gamma^* \leq U$. Thus,
$\gamma^* - L \leq U - L \leq \epsilon L$, or $\gamma^*  \leq (1 + \epsilon) L \leq (1 + \epsilon) \gamma$.

To prove the second claim, note that $\dens{S, \calG}$ is a rational number with a numerator of at most $n$. Thus, either $\gamma = \gamma^*$ or $\gamma^* - \gamma > n^{-2}$. If $\epsilon \leq \frac{1}{rn^3}$, then $\gamma^* - \gamma \leq U - L \leq n^{-2}$. Consequently, $\gamma = \gamma^*$.
\qed
\end{proof}

\algipcm requires $\bigO{\log rn - \log \epsilon}$ rounds, solving an integer linear program in each round. Note that solving an integer program is \np-hard~\cite{Schrijver1998}.

We should point out that fractional programming was used
by Goldberg~\cite{goldberg1984finding} for finding the densest subgraph in a single graph.
The key difference is the fairness constraint: without it $\problemcdcsm(\gamma)$ reduces to a minimum cut, which can be solved exactly in polynomial time.

\subsection{Exact algorithm for  \problemcdcsdiff}

Next, we will propose an exact algorithm to solve \problemcdcsdiff. The approach is similar to the solver for \problemcdcsm, that is, we will define an auxiliary problem, which can be solved with an integer program, and then use binary search to find the solution for \problemcdcsdiff. More formally, 
we have the following auxiliary problem  $\problemcdcsdiff(\gamma)$, and its relation to \problemcdcsdiff.

\begin{problem}[$\problemcdcsdiff(\gamma)$]
Given a graph sequence $\calG = (G_1,\dots,G_r)$, with  $G_i = (V, E_i)$, and two numbers $\alpha, \gamma$, find a subset of vertices $S \subseteq V$ minimizing  $b(S, \calG) - \gamma \abs{S}$, such that $m(S, \calG) \geq  \alpha\abs{S}$.
\end{problem}

\begin{proposition}
\label{prop:fracsbs}
Let $S(\gamma)$  be the subgraph  solving $\problemcdcsdiff(\gamma)$. Similarly, let $S^*$  be the optimal solution for \problemcdcsdiff. Write $\gamma^* =  \diff{S^*, \calG}$. If $\gamma < \gamma^*$, then $S(\gamma) = \emptyset$. If $\gamma > \gamma^*$, then $S(\gamma) \neq \emptyset$ and $\diff{S(\gamma),  \calG} < \gamma$.
\end{proposition}

\begin{proof}
Since an empty set satisfies the constraints, we have
$b(S(\gamma), \calG) - \gamma\abs{S(\gamma)} \leq 0$.
Let us first assume that $\gamma < \gamma^*$. If $S(\gamma) \neq \emptyset$, then $\diff{S(\gamma),  \calG} \leq \gamma < \gamma^*$ and $\dens{S(\gamma), \calG} \geq \alpha$, which contradicts the optimality of $\gamma^*$. Thus, $S(\gamma) = \emptyset$.

Next, let us assume that $\gamma > \gamma^*$. Then
\[
\begin{split}
    b(S(\gamma)) - \gamma \abs{S(\gamma)} & \leq b(S^*) - \gamma \abs{S^*}  < b(S^*) - \gamma^* \abs{S^*} = 0.
\end{split}
\]
Thus $b(S(\gamma)) <  \gamma \abs{S(\gamma)}$, consequently  $S(\gamma) \neq \emptyset$ and $\diff{S(\gamma), \calG} < \gamma$. \qed
\end{proof}
 
We can solve $\problemcdcsdiff(\gamma)$ with an integer linear program,

\begin{align*}
\textsc{minimize}&&  u - \ell & - \gamma \sum_{i = 1}^n y_i  \nonumber\\  
\textsc{subject to} && 
  x_{ij} & \leq y_i &  ij \in E   \\
&& x_{ij} & \leq y_j &  ij \in E \\
&& x_{ij} & \geq y_i + y_j - 1 &  ij \in E \\
&&  \sum_{ ij \in E_k} x_{ij} & \geq \ell &  k = 1,\dots, r \\
&&  \sum_{ ij \in E_k} x_{ij} & \leq u  &  k = 1,\dots, r \\
&& \sum_{k = 1}^r \sum_{ ij \in E_k} x_{ij} & \geq  \sigma  \sum_{i = 1}^{n} y_i\\
&& x_{ij}, y_j  & \in \{0,1\} \nonumber\\
&&  u,\ell & \geq 0 \nonumber\quad.
\end{align*}

Here we introduce upper and lower threshold variables $u$ and $\ell$ such that the difference $u - \ell$ matches to $b(S, \calG)$.
Similar to \algipcm, we search for the smallest $\gamma$ such that the solution is non-empty, that is, we run a binary search and stop when $U - L \leq (1 + \epsilon)L$, and then use $\problemcdcsdiff(U)$ as the final result. We refer to our algorithm as \algipdiff. The algorithm yields the following guarantees.

\begin{proposition}
Assume a graph sequence $\calG = (G_1,\dots,G_r)$, $\sigma > 0$, and $\epsilon > 0$. 
Let $\gamma$ be the score of the solution returned by $\algipdiff$ and let $\gamma^*$ be the optimal score of \problemcdcsdiff. Then $\gamma \leq (1 + \epsilon)\gamma^*$.
If $\epsilon \leq \frac{1}{n^3}$, then $\gamma = \gamma^*$.
\label{prop:opt-sbs-approx}
\end{proposition}
The proof is similar to the proof of Proposition~\ref{prop:opt-ip-approx}, and therefore is omitted.

\subsection{Greedy algorithm to find a good solution for  \problemcdcsdiff}

A standard technique in designing a polynomial-time algorithm is to start from the integer program, in this case \algipdiff, relax the integrality constraints, solve the resulting linear program, and then devise a rounding step. In our experimentation, this approach was problematic as the resulting sets rarely satisfied the constraints.
Consequently, we propose a greedy algorithm for \problemcdcsdiff. 

We start by solving \problemdts; let $S$ be the resulting set. Note since $\dens{S, \calG}$ is optimal,
$\dens{S, \calG} \geq \sigma$  or there is no set satisfying the constraint. 

We continue by trying to minimize $\diff{S, \calG}$ either by adding or removing a vertex greedily, while satisfying the density constraint. We repeat the same process until the algorithm
converges. The pseudo-code for the algorithm is given in Algorithm~\ref{alg:greedy}.

\begin{algorithm}[t!]
\caption{$\alggrd(\calG,\alpha)$, finds greedily a subgraph $S$ which minimizes $\diff{S, \calG}$ while satisfying $\dens{S, \calG} \geq \sigma$.}
\label{alg:greedy}
    $S \define $ The solution of \problemdts\;
	
    \While {changes to $\diff{S}$}{
        Find a vertex $v$ which minimizes $\diff{S}$ either by adding or removing while satisfying density constraint\;
        Add or remove $v$ from the set $S$\;    
    }
	\Return $S$\;
\end{algorithm}

\subsection{Greedy algorithm to find a good solution for  \problemcdcsm} \label{subsec:greedy-algo-fds}

Next, we present a two-phase greedy algorithm to find a good solution for \problemcdcsm. Given an input parameter $\alpha$, in the first phase of the algorithm, we search for a set $S$ such that $\diff{S, \calG} \leq \alpha$.

More concretely,  we repeatedly run \alggrd by varying $\sigma$ between $0$ and \dentds, the density of the solution for \problemdts, and choose a feasible set with the highest total density. As a candidate set for $\sigma$, we used $\set{\frac{i}{k}\dentds \mid i = 0, \ldots, k}$, where $k$ is a user parameter.

In the second phase of the algorithm, we start from our feasible set returned in the first phase, and at each iteration, we try to improve our density score by picking the best vertex to either add or delete until convergence. We refer to our algorithm as \alggrdfms.

\subsection{Exact algorithm for \problemdcs}

Although heuristics have been proposed to  \problemdcs problem~\cite{jethava2015finding, semertzidis2019finding}, we propose an integer programming-based algorithm to find an exact or near-optimal solution which we refer to this algorithm as \algipdcs. 

We introduce the following auxiliary problem.

\begin{problem}[$\problemdcs(\gamma)$]
Given a graph sequence $\calG = (G_1,\dots,G_r)$, with  $G_i = (V, E_i)$, find a common subset of vertices
$S$, 
such that $\min_i m(S, G_i) - \gamma \abs{S}$  is maximized.
\end{problem}

We solve $\problemdcs(\gamma)$ with the following integer program

\begin{align*}	\textsc{maximize\,\,\,\,}\hspace{.1in}&&\hspace{.05in}  z &- \gamma \sum_{i = 1}^{n} y_i 
	\\  
	\textsc{subject to}\hspace{.2in} && 
  x_{ij} &\leq y_i &  ij \in E \\
	&& x_{ij} &\leq y_j & ij \in E \\ 
 	&& \sum_{ i, j \in E_k} x_{ij} &\geq z  &  k = 1,\dots, r \\
	&& x_{ij}, y_j &\in \{0,1\}\\
        &&  z & \geq 0 \quad.
\end{align*}

To see why this program solves $\problemdcs(\gamma)$,
note that $x_{ij} = \min(y_i, y_j)$ and $z = \min_k m(S, G_k)$,
where $S = \set{i \mid y_i = 1}$.

We search for maximal $\gamma$ that induces non-empty solution
with a binary search. Here, we set the initial interval $(L, U)$ to $L = 0$ and $U =\frac{n - 1}{2}$, and keep halving the interval until $\abs{ U - L} \leq  \epsilon L$, where $\epsilon > 0$  is an input parameter. Finally, we return the solution of $\problemdcs(L)$ as the final solution to \problemdcs. 

The following result states the approximation guarantees of \algipdcs.

\begin{proposition}
Assume a graph sequence $\calG = (G_1,\dots,G_r)$ and $\epsilon > 0$. 
Let $\gamma$ be the score of the solution returned by \algipdcs and let $\gamma^*$ be the optimal score of \problemdcs. Then $\gamma \geq \gamma^*/(1 + \epsilon)$. If $\epsilon \leq \frac{1}{n^3}$, then $\gamma = \gamma^*$.
\label{prop:opt-dcs-approx}
\end{proposition}

The proof of the proposition is essentially the same as the proofs for \algipcm, and is therefore omitted.

\section{Related work}\label{sec:related}

Given an undirected graph, the problem of finding the densest subgraph can be solved in polynomial time. To solve the problem Goldberg~\cite{goldberg1984finding} proposed
an exact algorithm based on fractional programming, binary search, and minimum-cut computations. Moreover, Charikar~\cite{charikar2000greedy} proposed a linear program solving the problem, and a linear time, greedy $1/2$-approximation algorithm.

Instead of a single graph, the densest subgraph problem has been extended to the case of multiple graphs.
Jethava and Beerenwinkel~\cite{jethava2015finding} considered finding a common that maximizes the minimum density, an \np-hard problem as shown by Charikar et al.~\cite{charikar_on_finsing_dcs}.
We refer to this problem as \problemdcs problem. 
While LP-based and greedy algorithms proposed by Jethava and Beerenwinkel\cite{jethava2015finding} and the greedy-like algorithms proposed by Semertzidis et al.~\cite{semertzidis2019finding}  do not provide any theoretical approximation guarantees for all inputs, we propose an exact algorithm for \problemdcs problem as an additional result.

 Another problem variant for multiple graphs has been considered by Semertzidis et al.~\cite{semertzidis2019finding} where their objective was to maximize the sum of the densities. This problem is solvable in polynomial time by first flattening the graph sequence into one weighted graph and then considering it as the standard weighted densest subgraph instance. Arachchi and Tatti~\cite{arachchi2023jaccard} and Rozenshtein et al.~\cite{rozenshtein2020finding} considered extensions of the problem where the dense subgraph is allowed to vary across the snapshots.
 

Anagnostopoulos et al.~\cite{anagnostopoulos2020spectral} considered a single binary node-colored graph and constrained the subgraph to have a fair distribution of colors.
Even with two colors, the problem is shown to be \np-hard~\cite{anagnostopoulos2020spectral}.
Next, Miyauchi et al.~\cite{miyauchi2023densestdiverse} considered finding a fair densest subgraph of a node-colored graph with multiple colors.
Note that our problem can be viewed as finding fair densest subgraph in an edge-colored graph. Here, we color each edge with a color unique to a snapshot, and collapse the sequence into one (multi-edged) graph. Since the constraints and the input information in these problems are different the existing methods cannot be used to solve our problem.

Alternative measures for the density have been also studied. One
 option is to use the proportion of edges, that is, $m(S) / {\abs {S}  \choose 2}$. The issue with this measure is that a single edge yields the highest density of 1. Moreover, finding the largest graph with the edge proportion of 1 is equal to finding the largest clique, an inappproximable problem~\citep{hastad1996clique}.
 Tsourakakis et el~\cite{tsourakakis2013denser} proposed a score $m(S) - \alpha {\abs {S} \choose 2}$, leading to an \np-hard optimization problem. In addition, Tsourakakis~\cite{tsourakakis2015k} proposed a ratio of triangles 
 and the nodes as the density measure. Like the density based on edges, optimizing
 this measure can be done in polynomial time but requires computing the existing triangles in a graph. Adopting these measures with fairness constraints provides a future line of work.

\section{Experimental evaluation}\label{sec:exp}

The goal of this section is to experimentally evaluate our algorithms. To this end, we first generate a synthetic dataset with planted, fair dense component in each snapshot and test how well our algorithms discover the ground truth.
Next, we study the performance of the algorithm on real-world temporal datasets.
Finally, we present interpretative results from a case study.

We implemented the algorithms in Python\footnote{The source code is available at \url{https://version.helsinki.fi/dacs/}.
\label{foot:code}} and performed the experiments using a 2.4GHz Intel Core i5 processor and 16GB RAM. In our experimental evaluation, we used the Gurobi optimization solver in Python to solve the integer programs associated with \algipcm and \algipdiff algorithms by setting $\epsilon = 0.01$. As densities vary across the datasets,
for \problemcdcsdiff problem, we use $\sigma = \sbsnorm \times \dentds$ where $\dentds$ is the solution of \problemdts.
Recall that the greedy algorithm \alggrdfms requires finding a feasible set in the first phase, as discussed in Section~\ref{subsec:greedy-algo-fds}. The search is done with \alggrd and varying $\sbsnorm \in \set{\frac{i}{k} \mid i = 0, \ldots, k}$.
We initially set $k = 20$. If we did not find a feasible set, we tested $k = 100$. This strategy was sufficient for our experiments.

We compare our results with the exact solution of the densest common subgraph problem which maximizes $\min_i \dens{S, G_i}$~\cite{jethava2015finding}. We refer to this problem as \problemdcs.
We also compare our results to the total density induced by \problemdts~\cite{semertzidis2019finding}, and the sum of densities of individual dense subgraphs~\cite{goldberg1984finding}.

\subsection{Synthetic dataset}
The synthetic dataset was generated as follows.
The dataset consists of $4$ graphs given as $\{G_1,\ldots,G_4\}$ with $200$ vertices. We split the vertices into two groups $U$ and $W$.
For $G_1$ and $G_2$ we randomly placed $1500 p_i$ edges connecting $U$, where $p_i$ was sampled uniformly from the interval $[0.4, 1]$. We also randomly connected vertices $W$ with $200$ edges, as well as vertices $U$ and $W$ also with $200$ edges. For $G_3$ and $G_4$ we randomly connected vertices with $6000$ edges.
The resulting graph sequence had $14\,866$ edges. Moreover, the total density and the density difference for the dense fair component $U$ were $\dens{U, \calG} = 51.12$ and $\diff{U, \calG} = 3.9$.

\begin{table}[t!]
\setlength{\tabcolsep}{0pt}
\caption{
Computational statistics from the experiments with the synthetic datasets. Let $d_{\mathit{tds}}$ be the solution of \problemdts. Here,  \emph{constr.} relates to the input parameter $\alpha$ in \problemcdcsm or the normalized parameter $\sbsnorm$ in \problemcdcsdiff, $d_{\mathit{dis}}$ is the
discovered sum of densities, $d_{\mathit{min}}$, $d_{\mathit{max}}$, and $d_{a}$ are the minimum, maximum, and average density induced by the discovered subgraph over the graph sequence, respectively, $i$ gives the number of iterations of the algorithms, \emph{Jacc.} is the Jaccard index between the solution set and ground truth set, $\abs{S}$ provides the size of the solution, and \emph{time} gives the computational time in seconds.
 }

\label{tab:stats-syn}
\begin{filecontents*}{syn-exp.csv}
algorithm,alpha,sum_den,min_den,max_den,diff,avg_den,size,no_iterations,time,j_t
\algipcm,3.9,51.12,10.84,14.74,3.9,12.78,100,8,127.2170401,1
\alggrdfms,3.9,50.53061224,10.73469388,14.58163265,3.846938776,12.63265306,98,13,81.42657804,0.9603960396
\algipdiff,0.687743845,51.12,10.84,14.74,3.9,12.78,100,8,159.399714,1
\alggrd,0.687743845,51.16504854,10.59223301,15.02912621,4.436893204,12.79126214,103,97,8.707787037,0.9519230769
\end{filecontents*}
\pgfplotstabletypeset[
    begin table={\begin{tabular*}{\textwidth}},
    end table={\end{tabular*}},
    col sep=comma,
	columns = {algorithm,alpha,sum_den,min_den, max_den,avg_den,diff,no_iterations,size, j_t, time},
    columns/algorithm/.style={string type, column type={@{\extracolsep{\fill}}l}, column name=\emph{Algorithm}},
    columns/n/.style={fixed, set thousands separator={\,}, column type=r, column name=$\abs{V}$},
    columns/m/.style={fixed, set thousands separator={\,}, column type=r, column name=$Exp[\abs{E}]$},
    columns/time_0/.style={fixed, set thousands separator={\,}, column type=r, precision = 0, column name=$time$},
    columns/time_1/.style={fixed, set thousands separator={\,}, column type=r, column name=$t_2$},
    columns/time_2/.style={fixed, set thousands separator={\,}, column type=r, column name=$t_3$},
    columns/j_t/.style={fixed, set thousands separator={\,}, column type=r, column name=Jacc.},
    columns/alpha/.style={fixed, set thousands separator={\,}, column type=r, column name=constr.},
    columns/sum_den/.style={fixed,set thousands separator={\,}, column type=r, column name=$d_{dis}$},
    columns/diff/.style={fixed,set thousands separator={\,}, column type=r, column name=$\Delta$},
    columns/true_den/.style={fixed, set thousands separator={\,}, column type=r, column name=$d_{true}$}, 
    columns/dcs_den/.style={fixed, set thousands separator={\,}, column type=r, column name=$d_{\mathit{mds}}$}, 
    columns/dense_den/.style={fixed, set thousands separator={\,}, column type=r, column name=$d_{den}$}, 
    columns/min_den/.style={fixed, set thousands separator={\,}, column type=r, column name=$d_{\mathit{min}}$}, 
    columns/max_den/.style={fixed, set thousands separator={\,}, column type=r, column name=$d_{\mathit{max}}$}, 
    columns/axmax/.style={fixed, set thousands separator={\,}, column type=r, column name=$d_{\mathit{con}}$}, 
    columns/avg_den/.style={fixed, set thousands separator={\,}, column type=r, column name=$d_{a}$}, 
    columns/j1/.style={fixed, set thousands separator={\,}, column type=r, column name=$j_t$}, 
    columns/j2/.style={fixed, set thousands separator={\,}, column type=r, column name=$j_d$}, 
    columns/min_jac_dis_0/.style={fixed, set thousands separator={\,}, column type=r, column name=$J_{min}$}, 
    columns/time/.style={fixed, set thousands separator={\,}, column type=r, precision = 0, column name=time},
    columns/min_jac_dis_1/.style={fixed, set thousands separator={\,}, column type=r, column name=$j_{2}$}, 
    columns/min_jac_dis_2/.style={fixed, set thousands separator={\,}, column type=r, column name=$j_{3}$}, 
    columns/avg_jac_0/.style={fixed, set thousands separator={\,}, column type=r, column name=$\rho$}, 
    columns/avg_jac_1/.style={fixed, set thousands separator={\,}, column type=r, column name=$a_{2}$}, 
    columns/avg_jac_2/.style={fixed, set thousands separator={\,}, column type=r, column name=$a_{3}$}, 
    columns/no_iterations/.style={fixed, set thousands separator={\,}, column type=r, column name=$i$},
    columns/size/.style={fixed, set thousands separator={\,}, column type=r, column name=$\abs{S}$},
    columns/dis_d_1/.style={fixed,set thousands separator={\,}, column type=r, column name=$d_2$},
    columns/dis_d_2/.style={fixed,set thousands separator={\,}, column type=r, column name=$d_3$},
    columns/dis_obj_1/.style={fixed,set thousands separator={\,}, column type=r, column name=$o_2$},
    columns/dis_obj_2/.style={fixed,set thousands separator={\,}, column type=r, column name=$o_3$},
    every head row/.style={
		before row={\toprule},
			after row=\midrule},
    every last row/.style={after row=\bottomrule},
]
{syn-exp.csv}
\end{table}

\begin{table}[t!]
\setlength{\tabcolsep}{0pt}
\caption{
Characteristics of real-world datasets. Here, $n$ and  $m$ are the number of
vertices and edges, $r$  is the number of
snapshots, \dentds is the solution of \problemdts, $\Delta_{\mathit{tds}}$ is the difference between the maximum and minimum density of the \problemdts solution, $d_{\mathit{ind}}$
gives the sum of densities of individual densest subgraph from each graph snapshot, and  \denmds and $\Delta_{\mathit{mds}}$ give the total density and the difference between the maximum and minimum density of the \problemdcs solution, respectively. 
}

\label{tab:stats4}
\begin{filecontents*}{real-data.csv}
dataset,n,m_t,tau,m,dcs_den,dense_den,dcs_min,dcs_sum,delta,delta_tds
\dtname{Twitter-\#},806,1518,15,101.2,9.8,38.79534314,0.3333333333,5.5,0.2083333333,2
\dtname{Hospital},75,1885,5,377,30.28557777,41.67813165,2.823529412,17.76470588,2.470588235,6.928571429
\dtname{Airports},417,3588,37,96.97297297,24.53658537,83.75300416,0.3088235294,16.33823529,1.264705882,2.975609756
\dtname{Students},889,5329,122,43.68032787,26.31578947,118.0057357,0.08510638298,17.14893617,0.2553191489,0.6842105263
\dtname{Tumblr},1980,5812,89,65.30337079,55.83333333,103.988648,0.2307692308,43.61538462,0.7692307692,1.333333333
\dtname{Facebook},4117,8646,104,83.13461538,14,88.6484016,0.04705882353,5.752941176,0.07058823529,0.5
\dtname{Twitter-user},4605,10155,93,109.1935484,23,90.63006825,0.1063829787,12.80851064,0.170212766,0.5
\end{filecontents*}
\pgfplotstabletypeset[
    begin table={\begin{tabular*}{\textwidth}},
    end table={\end{tabular*}},
    col sep=comma,
	columns = {dataset,n,m_t,tau,dense_den,dcs_den, dcs_sum,delta_tds,delta},
    columns/dataset/.style={string type, column type={@{\extracolsep{\fill}}l}, column name=\emph{Data}},
    columns/n/.style={fixed, set thousands separator={\,}, column type=r, column name=$n$},
    columns/m_t/.style={fixed, set thousands separator={\,}, column type=r, column name=$m$},
    columns/tau/.style={fixed, set thousands separator={\,}, column type=r, column name=$r$},
    columns/time_0/.style={fixed, set thousands separator={\,}, column type=r, column name=$t_1$},
    columns/time_1/.style={fixed, set thousands separator={\,}, column type=r, column name=$t_2$},
    columns/time_2/.style={fixed, set thousands separator={\,}, column type=r, column name=$t_3$},
    columns/delta/.style={fixed, set thousands separator={\,}, column type=r, column name=$\Delta_{\mathit{mds}}$},
    columns/delta_tds/.style={fixed, set thousands separator={\,}, column type=r, column name=$\Delta_{\mathit{tds}}$},
    columns/true_den/.style={fixed, set thousands separator={\,}, column type=r, column name=$d_{true}$}, 
    columns/dcs_den/.style={fixed, set thousands separator={\,}, column type=r, column name=\dentds}, 
    columns/dcs_sum/.style={fixed, set thousands separator={\,}, column type=r, column name=\denmds}, 
    columns/dense_den/.style={fixed, set thousands separator={\,}, column type=r, column name=$d_{\mathit{ind}}$}, 
    columns/min_jac_dis_0/.style={fixed, set thousands separator={\,}, column type=r, column name=$j_{1}$}, 
    columns/min_jac_dis_1/.style={fixed, set thousands separator={\,}, column type=r, column name=$j_{2}$}, 
    columns/min_jac_dis_2/.style={fixed, set thousands separator={\,}, column type=r, column name=$j_{3}$}, 
    columns/avg_jac_0/.style={fixed, set thousands separator={\,}, column type=r, column name=$j_{1}$}, 
    columns/avg_jac_1/.style={fixed, set thousands separator={\,}, column type=r, column name=$j_{2}$}, 
    columns/avg_jac_2/.style={fixed, set thousands separator={\,}, column type=r, column name=$j_{3}$}, 
    columns/itr_0/.style={fixed, set thousands separator={\,}, column type=r, column name=$i_1$},
    columns/itr_1/.style={fixed, set thousands separator={\,}, column type=r, column name=$i_2$},
    columns/dis_d_0/.style={fixed,set thousands separator={\,}, column type=r, column name=$d_1$},
    columns/dis_d_1/.style={fixed,set thousands separator={\,}, column type=r, column name=$d_2$},
    columns/dis_d_2/.style={fixed,set thousands separator={\,}, column type=r, column name=$d_3$},
    columns/dis_obj_1/.style={fixed,set thousands separator={\,}, column type=r, column name=$o_2$},
    columns/dis_obj_2/.style={fixed,set thousands separator={\,}, column type=r, column name=$o_3$},
    every head row/.style={
		before row={\toprule},
			after row=\midrule},
    every last row/.style={after row=\bottomrule},
]
{real-data.csv}
\end{table}

\subsection{Results for the synthetic dataset} 
The densest subgraph, that is, the solution to \problemdts, contained the whole graph with the density $74.33$ and the density difference of $22.84$. We tested our algorithm by setting the constraints to match the fair dense component, that is, we set $\alpha = 3.9$ for \problemcdcsm
and $\sbsnorm = 0.69$ so that $\sigma = 51.12$ for \problemcdcsdiff.  We report our results in Table~\ref{tab:stats-syn}.

The Jaccard indices in Table~\ref{tab:stats-syn} indicate that the exact algorithms \algipcm and \algipdiff algorithms discovered the underlying fair dense subgraph.
Next, we see that both \alggrdfms and \alggrd achieved a Jaccard index of $0.95$ and $0.96$, respectively, while being faster than their exact counterparts. The exact algorithms yielded solutions with slightly better scores than the greedy algorithms.

Finally, solving \problemdcs yielded the fair dense component $U$ since $U$ for this data also had the largest minimum degree $\min_i \dens{S, G_i}$.

\begin{table}[ht!]
\setlength{\tabcolsep}{0pt}
\caption{
Computational statistics from the experiments with real-world datasets using \algipcm and \alggrdfms algorithms 
which are denoted as $\algexact$ and $\alggrdshort$, respectively. Here, $\alpha$ is the input parameter in \problemcdcsm problem which is the minimum value of the allowed induced density difference,
 $d_{\mathit{sum}}$ is the discovered sum of densities, $\Delta$ gives the difference between the minimum and maximum density induced by the solution set, $i$ gives the number of iterations, \emph{size} gives the size of the discovered subgraph, and \emph{time} gives the computational time in seconds.
}

\label{tab:real}
\begin{filecontents*}{real-sum-diff.csv}
dataset,alpha,d1,min_den,max_den,diff1,avg_den,size1,i1,time1,d2,min_den_g,max_den_g,diff2,avg_den_g,size2,i2,time_a,init_g,time2
\dtname{Twitter-\#},0.3,6.45,0.3,0.6,0.3,0.43,20,18,13.97,1.75,0,0.25,0.25,0.1166666667,4,3,0.09467220306,2.347235918,2.441908121
,0.5,7.583333333,0.25,0.75,0.5,0.5055555556,12,18,2.98,5,0,0.5,0.5,0.3333333333,2,1,0.01889204979,2.309831858,2.328723907
,0.7,7.923076923,0.2307692308,0.9230769231,0.6923076923,0.5282051282,13,18,3.90,5,0,0.5,0.5,0.3333333333,2,1,0.0227689743,2.162693262,2.185462236
\dtname{Hospital},0.3,12.21428571,2.214285714,2.5,0.2857142857,2.442857143,14,12,22.58,4.714285714,0.7857142857,1,0.2142857143,0.9428571429,14,2,0.01639819145,4.097306967,4.113705158
,0.5,13.85714286,2.5,3,0.5,2.771428571,14,12,22.50,5,0.6875,1.1875,0.5,1,16,5,0.03481817245,1.447252989,1.482071161
,0.7,14.53333333,2.466666667,3.133333333,0.6666666667,2.906666667,15,12,12.39,8.333333333,1.166666667,1.833333333,0.6666666667,1.666666667,18,5,0.03771805763,2.762027979,2.799746037
\dtname{Airports},0.3,14.95714286,0.2571428571,0.5571428571,0.3,0.4042471042,70,17,1006.11,6.3,0,0.3,0.3,0.1702702703,50,5,0.3809359074,24.93252993,25.31346583
,0.5,17.01851852,0.1851851852,0.6851851852,0.5,0.45995996,54,17,127.18,9.875,0,0.5,0.5,0.2668918919,40,8,0.5482370853,30.40279007,30.95102715
,0.7,18.42857143,0.08163265306,0.7755102041,0.693877551,0.4980694981,49,17,105.69,12.27272727,0,0.696969697,0.696969697,0.3316953317,33,1,0.0641169548,26.25217319,26.31629014
\dtname{Students},0.3,22.65,0.025,0.325,0.3,0.1856557377,40,19,28.00,17.83333333,0,0.2916666667,0.2916666667,0.1461748634,24,2,0.2655699253,102.0891719,102.3547418
,0.5,25.63888889,0,0.5,0.5,0.210154827,36,19,5.13,25.16666667,0,0.5,0.5,0.206284153,18,2,0.2163410187,102.2318382,102.4481792
,0.7,26.31578947,0,0.6842105263,0.6842105263,0.2157031924,19,19,4.39,26.31578947,0,0.6842105263,0.6842105263,0.2157031924,19,5,0.6277208328,92.0264802,92.65420103
\dtname{Tumblr},0.3,29.40740741,0.1851851852,0.4814814815,0.2962962963,0.3304203079,27,20,73.83,12.85714286,0,0.2857142857,0.2857142857,0.1444622793,7,2,0.2821612358,14.06188416,14.3440454
,0.5,38.55555556,0.1666666667,0.6666666667,0.5,0.4332084894,18,19,14.16,26,0,0.5,0.5,0.2921348315,8,1,0.1382431984,13.23060298,13.36884618
,0.7,45.92307692,0.1538461538,0.8461538462,0.6923076923,0.5159896283,13,19,7.06,37,0,0.6666666667,0.6666666667,0.4157303371,3,1,0.1440269947,13.05013466,13.19416165
\dtname{Facebook},0.3,11.14285714,0,0.2857142857,0.2857142857,0.1071428571,7,22,14.24,7,0,0.25,0.25,0.06730769231,4,1,0.2527418137,25.80105805,26.05379987
,0.5,14,0,0.5,0.5,0.1346153846,2,22,11.14,14,0,0.5,0.5,0.1346153846,2,1,0.2417669296,29.1269021,29.36866903
,0.7,14,0,0.5,0.5,0.1346153846,2,22,12.42,14,0,0.5,0.5,0.1346153846,2,1,0.264029026,29.34688711,29.61091614
\dtname{Twitter-user},0.3,23,0,0.5,0.5,0.247311828,4,21,15.14,11.5,0,0.25,0.25,0.123655914,4,1,0.3090670109,47.14920783,47.45827484
,0.5,23,0,0.5,0.5,0.247311828,4,21,14.32,23,0,0.5,0.5,0.247311828,2,1,0.3676872253,42.45010304,42.81779027
,0.7,23,0,0.5,0.5,0.247311828,4,21,14.32,23,0,0.5,0.5,0.247311828,2,1,0.317858696,44.59093189,44.90879059
\end{filecontents*}
\pgfplotstabletypeset[
    begin table={\begin{tabular*}{\textwidth}},
    end table={\end{tabular*}},
    col sep=comma,
	columns = {dataset,alpha,d1,d2,diff1,diff2, time1,time2,size1,size2,i1,i2},
    columns/dataset/.style={string type, column type={@{\extracolsep{\fill}}l}, column name=\emph{Data}},
    columns/time1/.style={fixed, set thousands separator={\,}, column type=r, precision = 0, column name=$\algexact$},
    columns/time2/.style={fixed, set thousands separator={\,}, column type=r, column name=$\alggrdshort$},
    columns/alpha/.style={fixed, set thousands separator={\,}, column type=r, column name=$\alpha$},
    columns/diff1/.style={fixed, set thousands separator={\,}, column type=r, column name=$\algexact$}, 
    columns/diff2/.style={fixed, set thousands separator={\,}, column type=r, column name=$\alggrdshort$}, 
    columns/axdts/.style={fixed, set thousands separator={\,}, column type=r, column name=$d_{\mathit{con}}$}, 
    columns/d1/.style={fixed, set thousands separator={\,}, column type=r, column name=$\algexact$}, 
    columns/d2/.style={fixed, set thousands separator={\,}, column type=r, column name=$\alggrdshort$}, 
    columns/i1/.style={fixed, set thousands separator={\,}, column type=r, column name=$i_{\algexact}$},
    columns/i2/.style={fixed, set thousands separator={\,}, column type=r, column name=$i_{\alggrdshort}$},
    columns/size1/.style={fixed, set thousands separator={\,}, column type=r, column name=$\algexact$},
    columns/size2/.style={fixed, set thousands separator={\,}, column type=r, column name=$\alggrdshort$},
    every head row/.style={
		before row={\toprule
		& & 
		\multicolumn{2}{l}{$d_{\mathit{sum}}$} &
		\multicolumn{2}{l}{$\Delta$ } &
             \multicolumn{2}{l}{time} &
             \multicolumn{2}{l}{size}  \\
		\cmidrule(r){3-4}
		\cmidrule(r){5-6}
		\cmidrule(r){7-8}
		\cmidrule(r){9-10}
  },
			after row=\midrule},
    every last row/.style={after row=\bottomrule},
]
{real-sum-diff.csv}
\end{table}

\begin{table}[ht!]
\setlength{\tabcolsep}{0pt}
\caption{
Computational statistics from the experiments with real-world datasets using using \algipdiff and \alggrd 
algorithms which are denoted as $\algexact$ and $\alggrdshort$, respectively. Here, $\sigma$ is the input parameter in \problemcdcsdiff where $\sigma = \sbsnorm \times \dentds$, $d_{\mathit{sum}}$ is the sum of densities induced by the discovered subgraph, $\Delta$ gives the difference between the minimum and maximum density, \emph{size} gives the size of the discovered subgraph, \emph{time} gives the computational time in seconds, and the columns $i_{\algexact}$ and $i_{\alggrdshort}$ give the number of iterations for respective algorithms.
}
\label{tab:real-diff}
\begin{filecontents*}{real-diff.csv}
dataset,alpha,axdts,diff1,d1,size1,time1,i1,diff2,d2,size2,time2,i2,,,,,,
\dtname{Twitter-\#},0.3,2.94,,,,,,0.3333333333,3.333333333,3,0.122,5,,,,,,
,0.5,4.9,0.05714285714,4.914285714,35,92.04912186,21,0.5,5,2,0.1071281433,4,,,,,,
,0.7,6.86,0.3571428571,7,14,25.979,18,0.75,7.25,4,0.108,4,,,,,,
\dtname{Hospital},0.3,9.085714286,0,9.666666667,15,297.395,34,1.583333333,9.125,24,0.193,21,,,,,0.189,21
,0.5,15.14285714,0.9333333333,15.2,15,32.095,13,2.733333333,15.16666667,30,0.139,13,,,,,0.142,13
,0.7,21.2,2.590909091,21.22727273,22,28.146,12,4.147058824,21.23529412,34,0.105,9,,,,,0.111,9
\dtname{Airports},0.3,7.36097561,,,,,,0.425,7.4,40,1.809,34,,,,,,
,0.5,12.26829268,,,,,,0.696969697,12.27272727,33,1.032,19,,,,,,
,0.7,17.17560976,0.5263157895,17.19298246,57,457.619,17,1.333333333,17.22222222,36,0.786,14,,,,,,
\dtname{Students},0.3,7.89473665,,,,,,0.125,7.925,40,3.722,30,,,,,,
,0.5,13.15789442,,,,,,0.2631578947,13.15789474,19,0.967,9,,,,,,
,0.7,18.42105218,0.171875,18.421875,64,555.533,19,0.3043478261,18.47826087,23,1.024,9,,,,,,
\dtname{Tumblr},0.3,16.75,,,,,,0.375,17.625,8,0.922,7,,,,,,
,0.5,27.91666666,0.2727272727,27.95454545,22,154.086,20,0.5714285714,28.28571429,7,0.477,4,,,,,,
,0.7,39.08333333,0.5263157895,39.15789474,19,61.396,19,0.8,39.6,5,0.267,2,,,,,,
\dtname{Facebook},0.3,4.2,,,,,,0.1428571429,4.285714286,7,1.689,6,,,,,,
,0.5,7,0.06666666667,7.016666667,60,961.776,23,0.25,7,4,0.843,3,,,,,,
,0.7,9.8,0.2142857143,9.857142857,14,43.187,21,0.3333333333,10,3,0.498,2,,,,,,
\dtname{Twitter-user},0.3,6.9,,,,,,0.1428571429,7.142857143,7,1.673,6,,,,,,
,0.5,11.5,,,,,,0.25,11.5,4,0.769,3,,,,,,
,0.7,16.1,0.1923076923,16.11538462,26,131.641,22,0.5,23,2,0.480,2,,,,,,
\end{filecontents*}
\pgfplotstabletypeset[
    begin table={\begin{tabular*}{\textwidth}},
    end table={\end{tabular*}},
    col sep=comma,
	columns = {dataset,alpha, axdts,diff1,diff2,d1,d2, time1,time2,size1,size2,i1,i2},
    columns/dataset/.style={string type, column type={@{\extracolsep{\fill}}l}, column name=\emph{Data},},
    columns/time1/.style={fixed, set thousands separator={\,}, column type=r, precision = 0, column name=$\algexact$},
    columns/time2/.style={fixed, set thousands separator={\,}, column type=r, column name=$\alggrdshort$},
    columns/alpha/.style={fixed, set thousands separator={\,}, column type=r, column name=$\sbsnorm$},
    columns/diff1/.style={fixed, set thousands separator={\,}, column type=r, column name=$\algexact$}, 
    columns/diff2/.style={fixed, set thousands separator={\,}, column type=r, column name=$\alggrdshort$}, 
    columns/axdts/.style={fixed, set thousands separator={\,}, column type=r, column name=$\sigma$}, 
    columns/d1/.style={fixed, set thousands separator={\,}, column type=r, column name=$\algexact$}, 
    columns/d2/.style={fixed, set thousands separator={\,}, column type=r, column name=$\alggrdshort$}, 
    columns/i1/.style={fixed, set thousands separator={\,}, column type=r, column name=$i_{\algexact}$},
    columns/i2/.style={fixed, set thousands separator={\,}, column type=r, column name=$i_{\alggrdshort}$},
    columns/size1/.style={fixed, set thousands separator={\,}, column type=r, column name=$\algexact$},
    columns/size2/.style={fixed, set thousands separator={\,}, column type=r, column name=$\alggrdshort$},
    every head row/.style={
		before row={\toprule
		& & &
		\multicolumn{2}{l}{$\Delta$} &
		\multicolumn{2}{l}{$d_{\mathit{sum}}$} &
             \multicolumn{2}{l}{time} &
             \multicolumn{2}{l}{size}  \\
		\cmidrule(r){4-5}
		\cmidrule(r){6-7}
		\cmidrule(r){8-9}
		\cmidrule(r){10-11}
  },
			after row=\midrule},
    every last row/.style={after row=\bottomrule},
]
{real-diff.csv}
\end{table}

\subsection{Real-world datasets} 
We consider $7$ publicly available, real-world datasets. The details of the datasets are shown in Table~\ref{tab:stats4}.
\dtname{Twitter-\#}~\cite{tsantarliotis2015topic}\footnote{\url{https://github.com/ksemer/BestFriendsForever-BFF-}\label{foot:gitbff}}  is a hashtag network where nodes correspond to hashtags and edges corresponds to the
interactions where two hashtags appear in a tweet. This dataset contains $15$ such daily graph snapshots in total. 
\dtname{Facebook}~\cite{viswanath2009evolution}\footnote{\url{https://networkrepository.com/fb-wosn-friends.php} } is a network of Facebook users in  New Orleans regional
community.  It contains a set of Facebook wall posts among these users from 9th of June to 20th of August, 2006.
\dtname{Students}\footnote{\url{https://toreopsahl.com/datasets/\#online_social_network}} is an online message
network at the University of California, Irvine. It spans over $122$ days.
\dtname{Twitter-user}~\cite{rozenshtein2020finding}\footnote{\url{https://github.com/polinapolina/segmentation-meets-densest-subgraph}} is a network of twitter users in Helsinki 2013.
It contains a set of tweets that appear in each others' names.
\dtname{Tumblr} ~\cite{leskovec2009meme}\footnote{\url{http://snap.stanford.edu/data/memetracker9.html}} contains 
phrase or quote mentions appeared in blogs and news media.  It contains author and meme interactions of users over $3$ months from February to April in 2009. Finally, the datasets, \dtname{Hospital} and \dtname{Airport}~\cite{oettershagen2024finding}\footnote{\url{https://gitlab.com/densest/diverse } \label{foot:diverse-repo}} represent multi-layer networks.

\subsection{Results for the real-world datasets}
First, let us compare the scores produced by the baselines \problemdts and \problemdcs, given in Table~\ref{tab:stats4}. By definition, the density $\dentds$ is always larger than $\denmds$ as \problemdts optimizes the total density. Moreover, $\Delta_{\mathit{tds}}$ is always greater than $\Delta_{\mathit{mds}}$, which is a side-effect of \problemdcs optimizing the minimum density.

Next, we report our results for the \problemcdcsm problem in Table~\ref{tab:real}. 
First, we see that the discovered scores $d_{\mathit{sum}}$ by both algorithms increase as $\alpha$ increases. This is because as we increase $\alpha$ we let the difference in the maximum and minimum density vary within a larger interval so that our objective score may increase. We see that $\Delta$ is close or equal to $\alpha$, and that \algipcm is almost always larger than \alggrdfms. This is expected since \algipcm searches for the densest subgraph more aggressively.

Next, we compare the scores and time of our exact and greedy algorithms. As expected, \algipcm achieves a higher score than \alggrdfms. 
We can see that \alggrdfms runs faster than \algipcm, but only for about half of the cases.
The reason for \alggrdfms being slower is the search for a feasible set as the algorithm uses \alggrd as a subroutine. Both algorithms required a low number of iterations.
The binary search in $\algipcm$ required solving at most $22$ integer programs. 
The number of iterations for $\alggrdfms$ is also low, at most $8$.

Next, we report our results for the \problemcdcsdiff problem in Table~\ref{tab:real-diff}.
Unlike with \algipcm,
in several cases, \algipdiff required an extensive amount of time: we stopped searching for the exact solution, if 
\algipdiff did not finish in $1$ hour. 
First, we see that the exact solution tends to have smaller densities $d_{\mathit{sum}}$ than the greedy since the exact algorithm optimizes $\diff{}$ more aggressively.

Next, we compare the scores and running times of our exact and greedy algorithms as shown in the $\Delta$ and time columns, respectively. As expected the objective increases as we tighten the density constraint by increasing $\sbsnorm$.
Moreover, \algipdiff achieves smaller scores than \alggrd at the cost of computational time. 
Finally, we find that the number of iterations is reasonable in practice with both algorithms as shown in $i_{\algexact}$ and $i_{\alggrdshort}$ columns.

\begin{figure}[t!]
\begin{subcaptiongroup}
\phantomcaption\label{fig:hu}
\phantomcaption\label{fig:hc}
\phantomcaption\label{fig:hu2}
\phantomcaption\label{fig:hc2}
\begin{center}

\setlength{\tabcolsep}{0pt}
\begin{tabular}{lll}
\begin{tikzpicture}[baseline = 0pt]
\begin{axis}[
    width=4.7cm,
    symbolic x coords={CIKM, WWW, SDM, PAKDD, ICDM, ECML, VLDB, WSDM, ICDE, KDD}, 
        ylabel = {Density},
        enlarge x limits={0.05},
        bar width = 7pt,
        xtick={CIKM, WWW, SDM, PAKDD, ICDM, ECML, VLDB, WSDM, ICDE, KDD},
        xtick style={draw=none},
        xticklabel style={rotate=45, font=\tiny, align=center, inner ysep=0pt, anchor=north east, yshift=2pt, xshift=2pt}]    \addplot[ybar,fill=yafcolor1,draw=none] coordinates {(CIKM,0)};
    \addplot[ybar,fill=yafcolor2,draw=none]coordinates {(WWW,1.75)};
    \addplot[ybar,fill=yafcolor3,draw=none]coordinates {(SDM,0)};
    \addplot[ybar,fill=yafcolor4,draw=none]coordinates {(PAKDD,0)};
    \addplot[ybar,fill=yafcolor5,draw=none]coordinates {(ICDM,0)};
    \addplot[ybar,fill=yafcolor6,draw=none]coordinates {(ECML,0)};
    \addplot[ybar,fill=yafcolor7,draw=none]coordinates {(VLDB,0)};
    \addplot[ybar,fill=yafcolor8,draw=none]coordinates {(WSDM,0)};
    \addplot[ybar,fill=yafcolor9,draw=none]coordinates {(ICDE,0)};
    \addplot[ybar,fill=yafcolor10,draw=none]coordinates {(KDD,0)};
\pgfplotsextra{\yafdrawyaxis{0}{1.75}}
\end{axis}
\node[anchor=north east] at (-0.5, -0.3) {(a)};
\end{tikzpicture} &
\begin{tikzpicture}[baseline = 0pt]
\begin{axis}[
    width=4.7cm,
    ymin = 0,
    ymax = 0.30,
    symbolic x coords={CIKM, WWW, SDM, PAKDD, ICDM, ECML, VLDB, WSDM, ICDE, KDD}, 
        enlarge x limits={0.05},
        bar width = 7pt,
        xtick={CIKM, WWW, SDM, PAKDD, ICDM, ECML, VLDB, WSDM, ICDE, KDD},
        xtick style={draw=none},
        xticklabel style={rotate=45, font=\tiny, align=center, inner ysep=0pt, anchor=north east, yshift=2pt, xshift=2pt}]    \addplot[ybar,fill=yafcolor1,draw=none] coordinates {(CIKM,0.2266666667)};
    \addplot[ybar,fill=yafcolor2,draw=none]coordinates {(WWW,0.2266666667)};
    \addplot[ybar,fill=yafcolor3,draw=none]coordinates {(SDM,0.1155555556)};
    \addplot[ybar,fill=yafcolor4,draw=none]coordinates {(PAKDD,0.06666666667)};
    \addplot[ybar,fill=yafcolor5,draw=none]coordinates {(ICDM,0.2133333333)};
    \addplot[ybar,fill=yafcolor6,draw=none]coordinates {(ECML,0.2266666667)};
    \addplot[ybar,fill=yafcolor7,draw=none]coordinates {(VLDB,0.05333333333)};
    \addplot[ybar,fill=yafcolor8,draw=none]coordinates {(WSDM,0.1822222222)};
    \addplot[ybar,fill=yafcolor9,draw=none]coordinates {(ICDE,0.02666666667)};
    \addplot[ybar,fill=yafcolor10,draw=none]coordinates {(KDD,0.02666666667)};  
\pgfplotsextra{\yafdrawyaxis{0}{0.3}}
\end{axis}
\node[anchor=north east] at (-0.2, -0.3) {(b)};
\end{tikzpicture} &
\begin{tikzpicture}[baseline = 0pt]
\begin{axis}[
    width=4.7cm,
    ymin = 0,
    ymax = 0.30,
    symbolic x coords={CIKM, WWW, SDM, PAKDD, ICDM, ECML, VLDB, WSDM, ICDE, KDD}, 
        xtick={CIKM, WWW, SDM, PAKDD, ICDM, ECML, VLDB, WSDM, ICDE, KDD},
        enlarge x limits={0.05},
        bar width = 7pt,
        xtick style={draw=none},
        xticklabel style={rotate=45, font=\tiny, align=center, inner ysep=0pt, anchor=north east, yshift=2pt, xshift=2pt}]
    \addplot[ybar,fill=yafcolor1,draw=none] coordinates {(CIKM,0.27307692307692305)};
    \addplot[ybar,fill=yafcolor2,draw=none]coordinates {(WWW,0.27307692307692305)};
    \addplot[ybar,fill=yafcolor3,draw=none]coordinates {(SDM,0.1)};
    \addplot[ybar,fill=yafcolor4,draw=none]coordinates {(PAKDD,0.057692307692307696)};
    \addplot[ybar,fill=yafcolor5,draw=none]coordinates {(ICDM,0.17307692307692307)};
    \addplot[ybar,fill=yafcolor6,draw=none]coordinates {(ECML,0.2692307692307692)};
    \addplot[ybar,fill=yafcolor7,draw=none]coordinates {(VLDB,0.06153846153846154)};
    \addplot[ybar,fill=yafcolor8,draw=none]coordinates {(WSDM,0.16538461538461538)};
    \addplot[ybar,fill=yafcolor9,draw=none]coordinates {(ICDE,0.019230769230769232)};
    \addplot[ybar,fill=yafcolor10,draw=none]coordinates {(KDD,0.007692307692307693)};
\pgfplotsextra{\yafdrawyaxis{0}{0.3}}
\end{axis}
\node[anchor=north east] at (-0.2, -0.3) {(c)};
\end{tikzpicture}\\
\end{tabular}
\end{center}
\end{subcaptiongroup}
\caption{Dense subgraphs among different conferences in \dtname{DBLP-CS}: (\ref{fig:hu}) \problemdts solution with density value of $1.75$, (\ref{fig:hc}) \problemcdcsm solution  with density value of $1.36$ for $\alpha = 0.2$, and (\ref{fig:hu2}) 
 \problemcdcsdiff solution with density value of $1.4$ for $\sigma = 0.8  \dentds$.
}
\label{fig:hist}
\end{figure}

\subsection{Case study}
\dtname{DBLP-CS}~\cite{oettershagen2024finding}\footref{foot:diverse-repo} is a co-authorship network where each edge represents a co-authorship attributed with one of the top data mining conferences. The graph contains $15\,308$  number of nodes and $10$ publication venues. We sampled  $12\,000$ edges from the original dataset.

We first solved \problemdts subgraph and observed that the densest subgraph represents only one conference which is the Web Conference (Fig.~\ref{fig:hu}). We are interested in finding the densest subgraph that contains publications of diverse publication venues without badly under-representing any venue. 
To that end, we applied $\algipcm(\alpha = 0.2)$ algorithm. In Figure~\ref{fig:hc} we see that the densest subgraph attained by \algipcm contains the edges from all publication venues in contrast to Figure~\ref{fig:hu}. The density of the unconstrained version was $1.75$ while our algorithm achieved a fair dense subgraph with $\diff{} = 0.2$ at the cost of having a density value of $1.36$.
Next, we applied $\algipdiff(\sigma = 0.8\dentds = 1.4)$ algorithm. As shown in Figure~\ref{fig:hc2}, it achieves a density value of $1.4$ while having with $\diff{} = 0.27$. 
Finally, we solve \problemdcs, yielding a solution with a density value of $1.18$ and $\diff{} = 0$. Note that \problemdcs does not maximize the density $\dens{S, \calG}$, and does not have a mechanism for controlling the trade-off between the $\diff{S, \calG}$ and $\dens{S, \calG}$. On the other hand, by changing the parameters $\sigma$ or $\alpha$ we can regulate the trade-off.

\section{Concluding remarks}\label{sec:conclusions}

In this paper, we introduced two variants of dense subgraph discovery problem for graphs with multiple snapshots that take fairness into account. 

More specifically,
given an input parameter $\alpha$, the goal of our first variant is to find a dense subgraph maximizing the sum of densities across snapshots such that the difference between the maximum and minimum induced density is at most $\alpha$. 
We considered also the dual problem where given an input parameter $\sigma$, we find a subgraph that minimizes the gap between the maximum and minimum density induced by the subgraph while inducing at least $\sigma$ amount of total density over the graph sequence.

We proved that both problems are \np-hard and proposed two exponential time, exact algorithms based on integer programming.
We also proposed two polynomial time heuristics.
We experimentally showed that our
algorithms could find the ground truth in synthetic dataset and perform reasonably well in real-world datasets.
Finally, we performed a study to show the usefulness of our problems.

The paper introduces several interesting directions for future work.  In this
paper, we considered the difference between the maximum and minimum density as the measure of fairness.
However, we can use an alternative constraint that forces the density induced in each snapshot to be at least a portion of the average density.
Another possible
direction is adopting a different kind of density definition for our problem setting.



\end{document}


\title{Fair densest subgraph across multiple graphs: supplementary material}



\maketitle


\appendix
\section{Solving minimum densest subgraph exactly}

As a baseline algorithm we consider the minimum densest subgraph problem~\cite{jethava2015finding} where the goal is to find a common subgraph which maximizes the minimum density.

\begin{problem}[Minimum densest subgraph~(\problemdcs)]
\label{pr:dcs}
Given a graph sequence $\calG = \set{G_1,\dots,G_r}$, with  $G_i = (V, E_i)$, find a common subset of vertices
$S$, 
such that $\min_i \dens{S, G_i}$, the minimum density induced by  $S$  over  $\calG$  is maximized.
\end{problem}
 
Although heuristics have been proposed to  \problemdcs problem~\cite{jethava2015finding, semertzidis2019finding}, we propose an integer programming-based algorithm to find an exact or near-optimal solution which we refer to this algorithm as \algipdcs. 

We use a similar approach as \algipcm algorithm where we utilize fractional programming and binary search. We introduce an auxiliary problem

\begin{problem}[$\problemdcs(\gamma)$]
Given a graph sequence $\calG = \set{G_1,\dots,G_r}$, with  $G_i = (V, E_i)$, find a common subset of vertices
$S$, 
such that $\min_i m(S, G_i) - \gamma \abs{S}$  is maximized.
\end{problem}

We solve $\problemdcs(\gamma)$ with the following integer program

\begin{align*}	\textsc{maximize\,\,\,\,}\hspace{.1in}&&\hspace{.05in}  z &- \gamma \sum_{i = 1}^{n} y_i 
	\\  
	\textsc{subject to}\hspace{.2in} && 
  x_{ij} &\leq y_i &  ij \in E \\
	&& x_{ij} &\leq y_j & ij \in E \\ 
 	&& \sum_{ i, j \in E_k} x_{ij} &\geq z  &  k = 1,\dots, r \\
	&& x_{ij}, y_j &\in \{0,1\}\\
        &&  z & \geq 0 \quad.
\end{align*}

To see why this program solves $\problemdcs(\gamma)$,
note that $x_{ij} = \min(y_i, y_j)$ and $z = \min_k m(S, G_k)$,
where $S = \set{i \mid y_i = 1}$.

We search for maximal $\gamma$ that induces non-empty solution
with a binary search. Here, we set the initial interval $(L, U)$ to $L = 0$ and $U =\frac{n - 1}{2}$, and keep halving the interval until $\abs{ U - L} \leq  \epsilon L$, where $\epsilon > 0$  is an input parameter. Finally, we return the solution of $\problemdcs(L)$ as the final solution to \problemdcs. 

The following result states the approximation guarantees of \algipdcs.

\begin{proposition}
Assume a graph sequence $\calG = \set{G_1,\dots,G_r}$ and $\epsilon > 0$. 
Let $\gamma$ be the score of the solution returned by \algipdcs and let $\gamma^*$ be the optimal score of \problemdcs. Then $\gamma \geq \gamma^*/(1 + \epsilon)$. If $\epsilon \leq \frac{1}{n^3}$, then $\gamma = \gamma^*$.
\label{prop:opt-dcs-approx}
\end{proposition}

The proof of the proposition is essentially the same as the proofs for \algipcm, and is therefore omitted.

\bibliographystyle{plainnat}
\bibliography{bibliography}